\documentclass[]{interact}

\usepackage{epstopdf}
\usepackage[caption=false]{subfig}
\usepackage{multirow}
\usepackage[numbers,sort&compress]{natbib}
\bibpunct[, ]{[}{]}{,}{n}{,}{,}
\renewcommand\bibfont{\fontsize{10}{12}\selectfont}
\makeatletter
\def\NAT@def@citea{\def\@citea{\NAT@separator}}
\makeatother

\theoremstyle{plain}

\theoremstyle{definition}

\theoremstyle{remark}

\begin{document}

\articletype{original article}

\title{Genetic Algorithm for quick finding of diatomic molecule potential parameters}

\author{
\name{Tomasz Urba\'nczyk and Jaros\l{}aw Koperski\thanks{CONTACT J.~K. Author. Email: ufkopers@cyf-kr.edu.pl}}
\affil{Smoluchowski Institute of Physcics, Jagiellonian University, \L{}ojasiewicza 11, 30-348 Krak\'{o}w, Poland}
}

\maketitle

\begin{abstract}
Application of Genetic Algorithm (GA) for determination of parameters of an analytical representation of diatomic molecule potential is presented. GA can be used for finding potential characteristics of an electronic energy state which can be described by analytical function. GA was tested on two artificially generated datasets which base on potentials with known characteristics and two LIF excitation spectra recorded using transitions in CdKr and CdAr molecules. Tests on generated datasets showed that GA can properly reproduce parameters of the potentials. Tests on experimental spectra indicated that changing the potential model from Morse, which is frequently used as a starting potential in IPA, to expanded Morse oscillator (EMO) leads to noticeable improvement of agreement between simulated and experimental data.	
\end{abstract}


\section{Introduction}

Inverse perturbation approach (IPA) \cite{KOSMAN_IPA, IPA_Asen} is a methodology which is widely used to obtain a pointwise potentials from spectroscopic data. The method is particularly useful for shallow or double-well potentials where other methods, \textit{e.g.} the Rydberg-Klein-Rees (RKR) procedure \cite{KIRSCHNER_RKR} do not work satisfactorily. In IPA method certain corrections to so-called starting potential are applied so, after solving the Schr{\"o}dinger equation, the calculated eigenvalues are close to the experimental energies of bound states. Sometimes, to find an appropriate representation of the potential, the process has to be repeated several times: the result of one IPA iteration determines the starting potential for new iteration. Therefore, the choice of a starting potential has a significant impact on lowering difficulties in employing IPA method. Using a correct starting potential - for which the eigenvalues are close to the real energies of the bound states - can greatly simplify the procedure \textit{i.e.}, by reducing the number of IPA iterations or by reducing vulnerability on the proper choice of IPA parameters. Often, for the starting potential, a Morse function is used with parameters obtained from so-called Birge-Sponer (B-S) fit to experimental data taking into account, that for IPA, the potential represented by a continuous function has to be converted into a pointwise form. Alternatively, \textit{ab-initio} calculated potential can be employed, however, its agreement with the experimental one is sometimes very limited.

In this article, we present a simple Genetic Algorithm (GA) for fitting an analytical potential to experimental data. GA can be used for finding potential characteristics in case of molecular electronic energy state that can be accurately represented by an analytical potential. Moreover, GA can be applied to generate a starting potential in IPA method, expecting higher accuracy than whilst it is represented with a Morse function obtained from B-S plot.

GA \cite{Mitchell_1998} is a heuristic procedure inspired by the Darwin’s theory of evolution. It can be used for searching solutions in optimization problems. GAs are used in a wide area of science and engineering \textit{e.g.}, in characterization of economic models \cite{ARIFOVIC_Economy_COBWEB}, planning trajectories for robot manipulators \cite{RobotTrajectoryPlanning} or designing vehicles \cite{VehicleDesigning}. They are also used in molecular spectroscopy. Roncaratti \textit{et al.} presented GA for fitting analytical potentials (Rydberg form) to \textit{ab initio} points for H$_2^+$ and Li$_2$ systems \cite{RONCARATTI}. Marques \textit{et al.} designed GA for direct fit of spectroscopic data which was successfully used for NaLi and Ar$_2$ \cite{Marques_2008}. Almeida \textit{et al.} expanded this method for very challenging potential of the RbCs ground state \cite{Almeida_2011}. Recently, Stevenson and P{\'{e}}rez-R{\'{\i}}o developed GA for fitting pointwise potentials to experimental data for diatomic molecules achieving 0.03 cm$^{-1}$ overall accuracy for the $ X^1\Sigma^+$ state in LiRb \cite{Stevenson_2019}. GAs are also used for analyzing spectra of large molecules. Meerts and Schmitt presented automated assignment and fitting procedure for high-resolution rotationally resolved spectra \cite{Meerts_IRPC}. They used the procedure \textit{e.g.}, on 4-methylphenol, resorcinol or benzonitrile and phenol dimers.

\section{Genetic Algorithms}

GA is an optimization algorithm inspired by biological evolution, especially by the natural selection process. To implement GA one should take into account three biological phenomena associated with reproduction of living organisms: selection, recombination (crossover) and mutation.

Let us assume a set of solutions (so-called candidate solutions) to a given optimization problem. The set, which initially can be generated randomly, is called population or generation. To implement selection it must exist so-called fitness function that quantitatively assesses the correctness of each candidate solution. The main goal of the selection is to pick these solutions which will be allowed to the reproduction process for forming the new generation of candidate solutions. To do this, solutions in current generation are ordered using the fitness function. After ordering, the more optimal solutions have higher positions on the list of solutions than their less optimal counterparts. According to evolution theory, the individuals which are better suited to the environment have higher chances for reproduction, so they traits can be more likely passed to the new generations than those of worse adapted population members. The same concept is used in GA. The higher position of particular candidate solution on the ordered list, the higher probability of involvement of this solution in the reproduction process. In the process, two candidate solutions (parents), are picked randomly from the current generation, taking into account that picking solutions with higher positions on the ordered list should be preferred. To produce an offspring, their parameters are combined together, through analogy to the biological genetic recombination process, \textit{e.g.} the parameters of the offspring can be chosen as an average or a weighted average of the parameters of its parents.

The last biological phenomenon which should be implemented in GA is mutation which helps to maintain diversity of solutions in subsequent generations. This is important, because if solutions in considered generation are too similar, the evolution slows. Moreover, in this situation algorithm may stuck near the local optimum. The implementation of mutation in GA can be realized by multiplying all parameters of the offspring solution by random numbers close to one. The range from which the random numbers are picked determines the strength of the mutation.

To simulate the evolution, the process of creation of subsequent generations is iteratively repeated. The generation sizes \textit{i.e.}, the numbers of candidate solutions in generations, are fixed in advance. The algorithm can be terminated if any of found solutions satisfy the minimum criterion (\textit{i.e.} the result of its fitness function exceed predefined value) or if the given maximum number of generations is reached.

On should notice, that due to the rules of mutation and recombination processes, evaluation of the fitness function applied to the offspring of two very good candidate solutions can return a very poor result. Especially, result of the fitness function of the offspring can be significantly worse than the results of its parents. This can lead to loss of very good candidate solutions from current generation during creation of new generation. To prevent this undesired situation and to guarantee that the quality of the best solution does not decrease from one generation to the next, GA should implement a concept of \textit{elitism}. According to this idea, a selected number of candidate solutions with the highest position on the ordered list of solutions should be transferred without any alteration to the new generation.

\subsection{Implementation of GA for determination of molecular potential parameters}
\begin{figure}[bt!]
	\centering
	\includegraphics[width=1.0\textwidth]{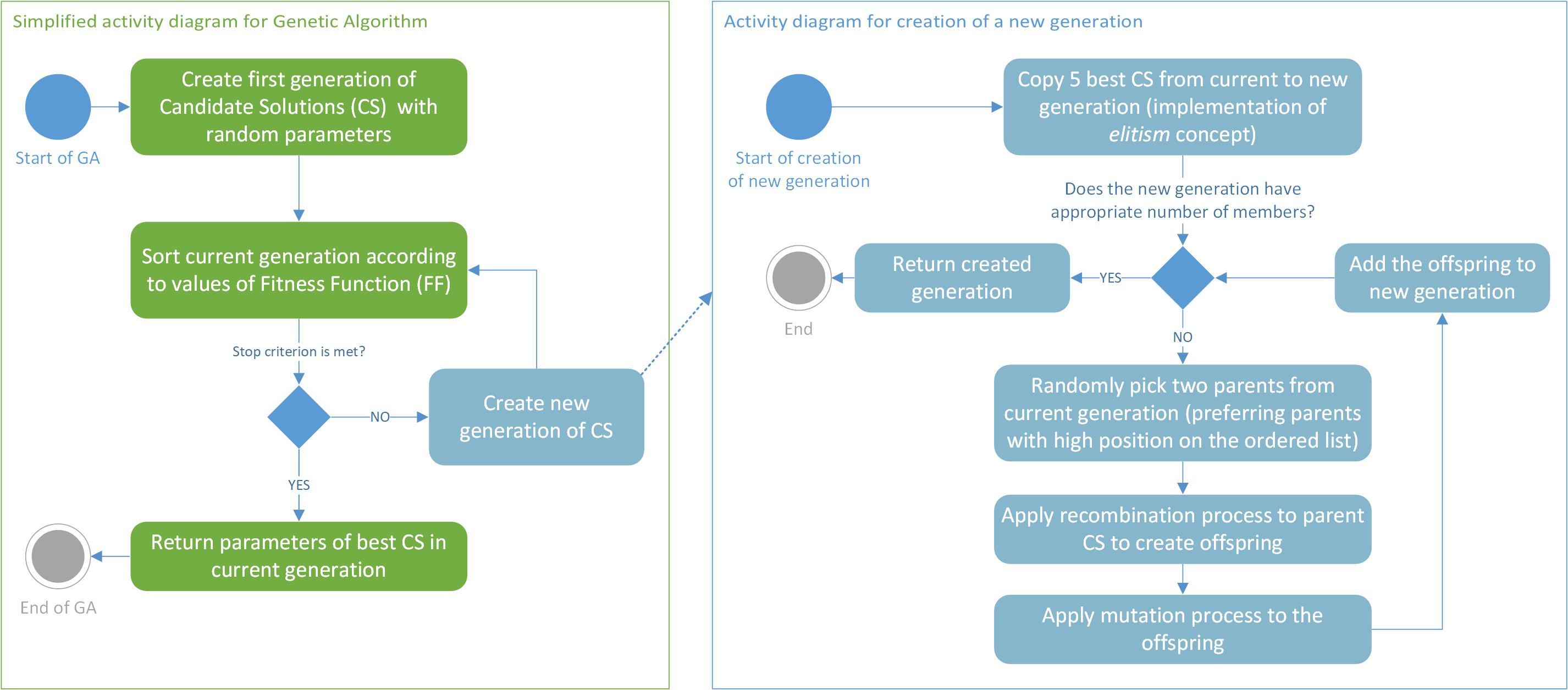}
	\caption{Left part: simplified activity diagram for Genetic Algorithm. The "Stop criterion" is set by the maximum number of generations. Right part: detailed activity diagram for creation of a new generation of candidate solutions.}
	\label{schemeGA}
\end{figure}

Our goal is to construct GA which can find parameters of given analytical molecular potential that results in simulation of energies of $(\upsilon,J)$ ro-vibrational levels being close to the experimental values. The simplified activity diagram for the algorithm is presented in Fig. \ref{schemeGA}.

Let us assume the interatomic potential of diatomic molecule given by a function $U=U(a_0,a_1...a_n)$, where $a_0,...,a_n$ are parameters. For a given potential characterized by the set of parameter values, which forms the candidate solution to the optimization problem, one can easily find simulated energies of $(\upsilon,J)$ levels $E_{\upsilon,J}^{sim}$ by solving appropriate Schr{\"o}dinger equation. To do this, we use LEVEL program \cite{LEVEL}, but different programs \textit{e.g.}, Duo \cite{DUO_Yourchenko} could be used as well. The quality of each solution is evaluated by a comparison of $E_{\upsilon,J}^{sim}$ with the experimental (or referenced) values $E_{\upsilon,J}^{exp}$. Here, we propose two representations of the so-called fitness function ($FF$). The first version of $FF$ is defined as

\begin{equation}\label{eqFF}
FF_{V1}=\sum_{\upsilon,J}|(E_{\upsilon,J}^{exp}-E_{0,0}^{exp})-(E_{\upsilon,J}^{sim}-E_{0,0}^{sim})|,
\end{equation}
where $E_{0,0}^{exp}$ and  $E_{0,0}^{sim}$ are experimental and simulated energies of one specified $(\upsilon,J)$ level \textit{e.g.}, $(\upsilon=0,J=0)$. In Eq. \ref{eqFF}, using differences between $E_{\upsilon,J}$ and $E_{0,0}$ comes from the fact, that from the experimental spectrum it is often easier and more reliable to obtain differences between energies of bound states rather than their absolute values. Alternatively, one can use second representation of the fitness function which relies on absolute values of the experimental energies

\begin{equation}\label{eqFFV2}
FF_{V2}=\sum_{\upsilon,J}|(E_{\upsilon,J}^{exp}-E_{\upsilon,J}^{sim})|.
\end{equation}

Here, we present results of tests for both versions of $FF$. For $FF_{V1}$ and $FF_{V2}$, the sum includes all $(\upsilon,J)$ levels present in the experimental spectrum. For a good candidate solution, FF returns low value, whereas for a bad solution high value is returned. It is also obvious, that for the ideal solution the sum is equal to zero. To sort the candidate solutions, we order them ascending with respect to the values returned by $FF$.

For recombination the process of random selection of candidate solutions was implemented using a random number generator. Firstly, we pick a random real number $x$ from the Gaussian distribution with distribution mean set to zero and standard deviation $\sigma$ set to one. Next, we find the index $i$ of the chosen candidate solution on the ordered list by formula

\begin{equation}
i=Floor(|A\cdot L\cdot x|),
\end{equation}
where $Floor$ is a function which takes the integer part from a real number, $L$ is length of the list of candidate solutions (\textit{i.e.}, number of solutions in the current generation) and $A$ is a chosen multiplying factor that is a hyperparameter of GA which in our implementation is set to 0.02 by default. If the calculated index $i$ is larger than $L-1$ (it is possible only for high values of $A$ when $|x|\geq1$) the process is repeated. To create a new offspring solution, two parents are selected independently. However, it is allowed that the offspring solution has the same candidate solution as both parents because - due to the mutation process - the offspring solution will have slightly different parameters than the "doubled" parent solution.

To implement the recombination (crossover) process, we averaged the parameters of both parents with randomly generated weights. Assuming that  $a_i^I$ and $a_i^{II}$ are $i-th$ parameters of first and second parent solution candidate, respectively, the offspring parameter $a_i^{off}$ is calculated according to the equation

\begin{equation}\label{recombEq} 
a_i^{off}=p\cdot a_i^I+(1-p)\cdot a_i^{II},
\end{equation}
where $p$ is a random real number from uniform distribution in the range from 0 to 1. In our implementation of GA, the new value of $p$ is generated for each parameter independently. To implement mutation, the result of Eq. \ref{recombEq} is multiplied by a random real number close to one

\begin{equation} \label{eqMut}
a_i^{off\,mut}=q\cdot a_i^{off}.
\end{equation}

In Eq. \ref{eqMut}, $q$  is a random real number from uniform distribution ranging from 1-$\epsilon$ to 1+$\epsilon$, where $\epsilon$ is the hyperparameter set to 0.005 by default.
The parameters of candidate solutions in first generation are picked randomly from ranges specified by the user. The common size of each generation is also specified by the user (usually, there is several hundred candidate solutions in one generation), however, we assume that the size of first generation is tripled comparing to the common size of other generations. To implement \textit{elitism} at the beginning of creation of the new generation, the algorithm copies specified number (5 by default) of the best candidate solutions from the current generation to the new one. The algorithm terminates after creating specified number of generations.

Presented implementation of GA was created in C\# language using Microsoft Visual Studio integrated development environment. The values of hyperparameters $A$ and $\epsilon$ were chosen arbitrarily after several trials. The selected values provided best performance of GA. It means that in the test, for selected values of $A$ and $\epsilon$, GA returns solution with small $FF$ after small number of generations.

\section{Tests on generated datasets}
To evaluate the correctness of our algorithm, we tested it on artificially generated reference datasets. These datasets contained simulated energies of $(\upsilon,J)$ levels, associated with simulations based on the interatomic potentials with known characteristics. Thanks to this approach, we can check if the parameters of the potential returned by GA are similar to parameters of the potential which was used to create reference data. The datasets were loosely inspired by the excitation spectrum of the $b^30_u^+(5^3P_1)\leftarrow X^10_g^+(5^1S_0)$ transition in Cd$_2$ \cite{Cd2IRPC}. We assumed that the artificial reference spectra contains first 15 vibrational components with 10 resolved rotational lines in each component. The datasets were generated under assumption that the potential of the $b^30_u^+$ state was expressed by expanded Morse oscillator (EMO) function, proposed by Le Roy and co-workers \cite{EMO1}

\begin{equation}
U(r)=D_e\left[1-e^{-\beta(r)\left(r-R_e\right)} \right]^2,  \,\,\,\,\,\, \beta(r)=\sum_{i=0}^{N} \beta_i \left(\frac{r-R_e}{r+R_e}\right)^i.
\end{equation}

GA is a heuristic method, therefore its result always contains some kind of randomness. Therefore, in each of the performed tests, GA was executed 15 times \textit{i.e.}, in each execution the algorithm had the same parameters. For all tests, the hyperparameters of GA were set to default values: $\epsilon=0.005$, $A=0.02$, while 5 candidate solutions were transferred to the new generation as a realization of the \textit{elitism} concept. Tests were performed independently for both proposed representation of $FF$: $FF_{V1}$ and $FF_{V2}$. We emphasize that GA can work with any analytical potential \textit{e.g.}, Lennard-Jones or double exponential long range (DELR) \cite{DELR} potentials. The advantage of EMO potential is that, it is an extension of a Morse potential, so we can relatively easy predict the searching ranges of its parameters (for EMO, $R_e$, $D_e$ and $\beta_0$  should be similar to the values used for a Morse function).

\subsection{Spectra without noise}
\begin{figure}[bt!]
	\centering
	\includegraphics[width=0.95\textwidth]{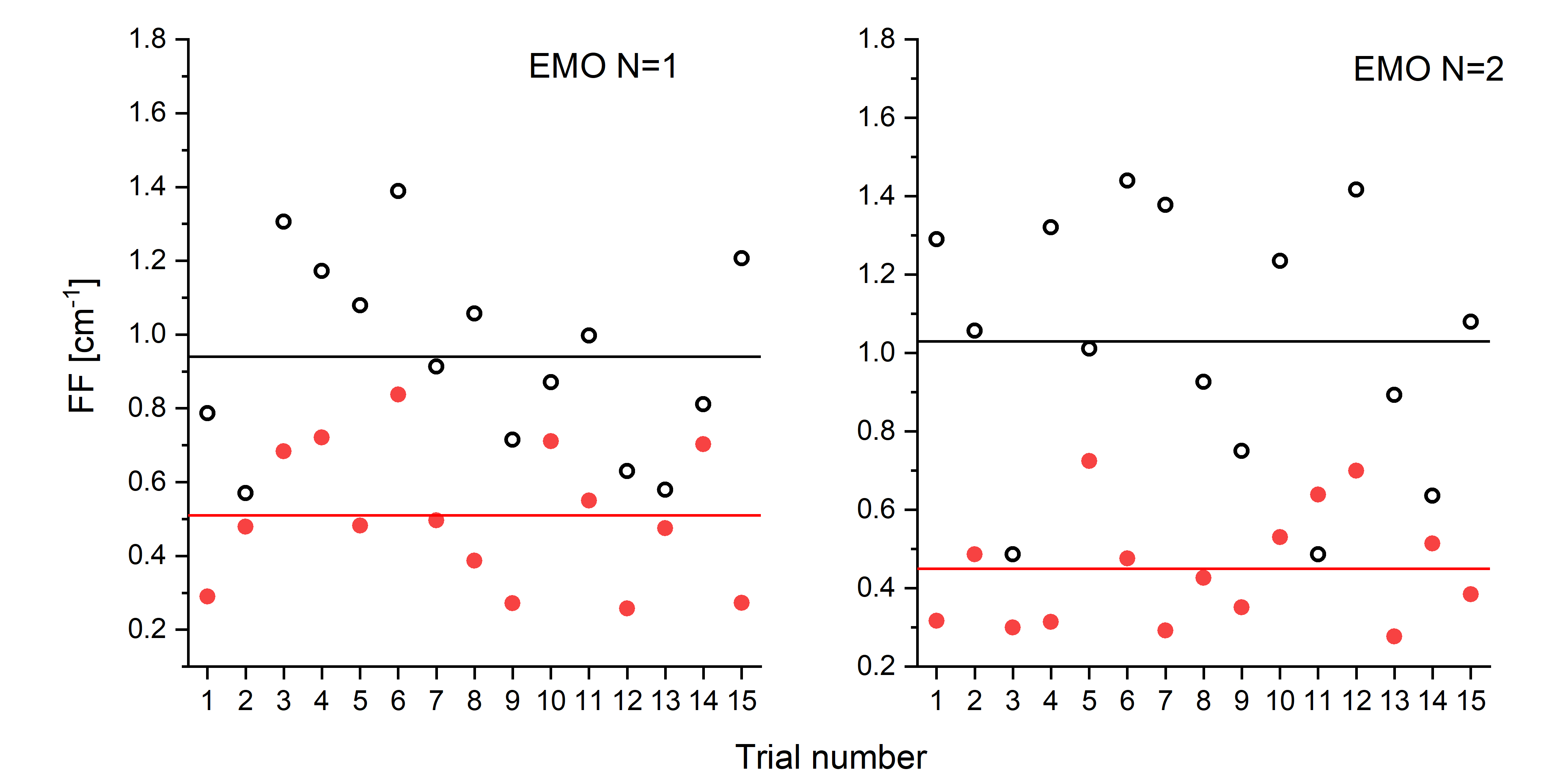}
	\caption{Fitness function ($FF$) obtained as a result of GA for the best candidate solution in 15 independent trials for EMO $N=1$ (left side) and EMO $N=2$ (right side).  $FF_{V1}$ and $FF_{V2}$ are represented with full red and black empty circles, respectively. The averages for $FF_{V1}$ and $FF_{V2}$ are drawn with horizontal red and black lines, respectively. Details in text.}
	\label{FigWithoutNoise}
\end{figure}

In the first test, we checked GA on two artificially generated datasets based on EMO potential with $N=1$ and $N=2$, respectively. In both tests, we terminated the algorithm after 15 generations, each generation having 800 candidate solutions, except first generations which had 2400 candidate solutions. The results of the tests are presented in Fig. \ref{FigWithoutNoise} and  Tables \ref{tabTest1} and \ref{tabTest2}.

\begin{table}[bt!]
	\footnotesize
	\caption{Results of searching of EMO $(N=1)$ potential parameters using GA for referenced data inspired by potential of the $b^30_u^+(5^3P_1)$ state in Cd$_2$ \cite{Cd2IRPC}.}
	\label{tabTest1}
	\begin{center}
		\begin{tabular}{cccccc}
			\hline\\[-8pt]
			Parameter&Searching&GA value$^a$&GA value$^a$&Expected\\ 
			&range&for $FF_{V1}$&for $FF_{V2}$&value$^b$\\ %
			\hline\\[-8pt]
			$R_e$[\AA] & 3.95-4.05 &4.020  &4.015 &4.020 \\
			$D_e[cm^{-1}]$ & 250-270 &259.56 &259.49 &259.50  \\
			$\beta_0$[1/\AA] & 0.8-1.4 &1.150 &1.150 &1.150\\
			$\beta_1$[1/\AA] & 0.0-0.6&0.177 &0.179 &0.180 \\	
			\hline\\[-8pt]
			${FF_{avg}}^c[cm^{-1}]$& &0.51(19) &0.94(26) & \\
			\hline\\[-8pt]
		\end{tabular}	
	\end{center}    
	$^a$ Result of one of 15 trials with lowest value of $FF$ (compare Fig. \ref{FigWithoutNoise}). Values determined using GA with 15 generation and 800 candidate solutions in the generation.  $FF$ for best solution returned 0.26 cm$^{-1}$ and 0.57 cm$^{-1}$ for $FF_{V1}$ and $FF_{V2}$, respectively. Details in text.\\	
	$^b$Parameters used for creation of the reference dataset.\\
	$^c$The average $FF$ for best solutions obtained in 15 trials (compare horizontal lines in Fig. \ref{FigWithoutNoise}),  the uncertainty obtained as $\sigma$.\\ 
\end{table}

\begin{table}[bt!]
	\footnotesize
	\caption{Results of searching of five EMO $(N=2)$ potential parameters using GA for referenced data inspired by the potential of the $b^30_u^+(5^3P_1)$  state in Cd$_2$ \cite{Cd2IRPC}.}
	\label{tabTest2}
	\begin{center}
		\begin{tabular}{ccccc}
			\hline\\[-8pt]
			Parameter&Searching&GA value$^a$&GA value$^a$&Expected\\ 
			&range&for $FF_{V1}$&for $FF_{V2}$&value$^b$\\ %
			\hline\\[-8pt]
			$R_e$[\AA] &3.95-4.05&4.007   &4.008  &4.010 \\
			$D_e[cm^{-1}]$ &250-270&257.05  &257.01 & 257.00 \\
			$\beta_0$[1/\AA] &0.8-1.4&1.100  & 1.100&1.100 \\
			$\beta_1$[1/\AA] &0-0.6&0.244  &0.248&0.250  \\	
			$\beta_2$[1/\AA] &0-0.25&0.167 &0.150&0.150 \\
			\hline\\[-8pt]
			${FF_{avg}}^c[cm^{-1}]$& &0.45(15) &1.03(33) & \\
			\hline\\[-8pt]
			
		\end{tabular}	
	\end{center}
	$^a$Result of one of 15 trials with lowest value of $FF$ (compare Fig. \ref{FigWithoutNoise}). Values determined by GA with 15 generation and 800 candidate solutions in the generation. $FF$ for best solution returned 0.28 cm$^{-1}$ and 0.49 cm$^{-1}$ for $FF_{V1}$ and $FF_{V2}$, respectively. Details in text.\\
	$^b$Parameters used for creation of the reference dataset.\\
	$^c$The average $FF$ for best solutions obtained in 15 trials (compare horizontal lines in Fig. \ref{FigWithoutNoise}),  the uncertainty obtained as $\sigma$.\\
\end{table}

For both EMO ($N=1$) and ($N=2$) potentials, the comparison shows a high degree of  agreement between parameters of the potential obtained using GA and parameters which were used to create the reference  $E_{\upsilon,J}^{exp}$ data.

From Fig. \ref{FigWithoutNoise} one can see, that for both EMO $(N=1)$ and $(N=2)$ potential representations, values of $FF$ for $FF_{V1}$ are smaller (on average) than that obtained for $FF_{V2}$. For EMO $(N=1)$ dataset, the averages of $FF_{V1}$ and $FF_{V2}$ are equal 0.51  $cm^{-1}$ and 1.00  $cm^{-1}$, respectively, whereas for EMO $(N=2)$, the averages of $FF_{V1}$ and $FF_{V2}$ take values 0.45  $cm^{-1}$ and 1.03  $cm^{-1}$, respectively. $FF$ measures the sum of discrepancies between simulated and reference $(\upsilon,J)$ levels. For both EMO representation in all trials, $FF_{V1}$ for the best candidate solution took value less than 1.0  $cm^{-1}$ (0.006  $cm^{-1}$ per $(\upsilon,J)$ level), whereas $FF_{V2}$ was less than 1.8  $cm^{-1}$ (0.010  $cm^{-1}$ per $(\upsilon,J)$ level). Obtained results indicate that GA can be used to obtain interatomic potentials which correctly reproduce reference or experimental energies of $(\upsilon,J)$ levels.

\subsection{Performance of GA for "noised" experimental data}
\label{noised}
The experimental energies of $(\upsilon,J)$ levels are always measured with some uncertainty. To test the performance of GA in case of "noised" experimental data, we artificially generated dataset for EMO $(N = 1)$ potential containing 176 $(\upsilon,J)$ levels with added random noise. It is necessary to emphasize, that in case of $FF_{V2}$, GA works directly on energies of $(\upsilon,J)$ levels from the dataset, while in case of $FF_{V1}$, GA uses differences between energies of $(\upsilon,J)$ levels in the dataset and the energy of "reference" $(\upsilon,J)$ level: $(\upsilon=0,J=0)$. Adding noise to the dataset for $FF_{V2}$ is straightforward. To do this, to each energy of $(\upsilon,J)$  level we add randomly chosen real number $N$ from a Gaussian distribution with the mean set to 0 cm$^{-1}$ and the selected $\sigma$, where the value of $\sigma$ is connected with the "strength" of the noise.

However, to include noise to the dataset for $FF_{V1}$ we added random noise to energy of each \textit{difference} $E_{\upsilon,J}^{exp}-E_{0,0}^{exp}$ considered in the $FF$. Therefore, we added random noise to each $(\upsilon,J)$ level except the "reference" level $(\upsilon=0,J=0)$. The noise was added in the same manner as for $FF_{V2}$. The level $(\upsilon=0,J=0)$ is free of noise due to the fact that addition of a random noise to its energy would change the problem which is solved by GA. Please notice, that if a random noise is added to all $(\upsilon,J)$ levels, except $(\upsilon=0,J=0)$, the average $\overline{(E^{exp\,rand}_{\upsilon,J}-E^{exp}_{0,0})}$ will be the same as the average calculated for data without noise: $\overline{(E^{exp}_{\upsilon,J}-E^{exp}_{0,0})}$. In the expression above, the $E^{exp\,rand}_{\upsilon,J}$ denotes the energies of ($\upsilon,J$) levels from the reference dataset with added noise. If a random noise is added to all levels, the average for "noised" data will vary from the average for data without the noise by the noise added to $(\upsilon=0,J=0)$. If, in case of using $FF_{V1}$, a random noise is added also to the $E^{exp}_{0,0}$, the solution found by GA is stable, but obtained potential parameters can be different from these used to create reference dataset. However, if the noise is added to all ($\upsilon,J$) levels, the potential found by GA can properly simulate differences between energy levels in the prepared "noised" dataset. In case of real experimental data, especially in cases when some parts of the spectrum have better signal-to-noise ratio (SNR) than other parts, instead of using in $FF_{V1}$ the energy of lowest $(\upsilon,J)$ level $E^{exp}_{0,0}$, one can employ the energy of different $(\upsilon,J)$ level that is determined with the highest precision.

\begin{figure}[bt!]
	\centering
	\includegraphics[width=0.8\textwidth]{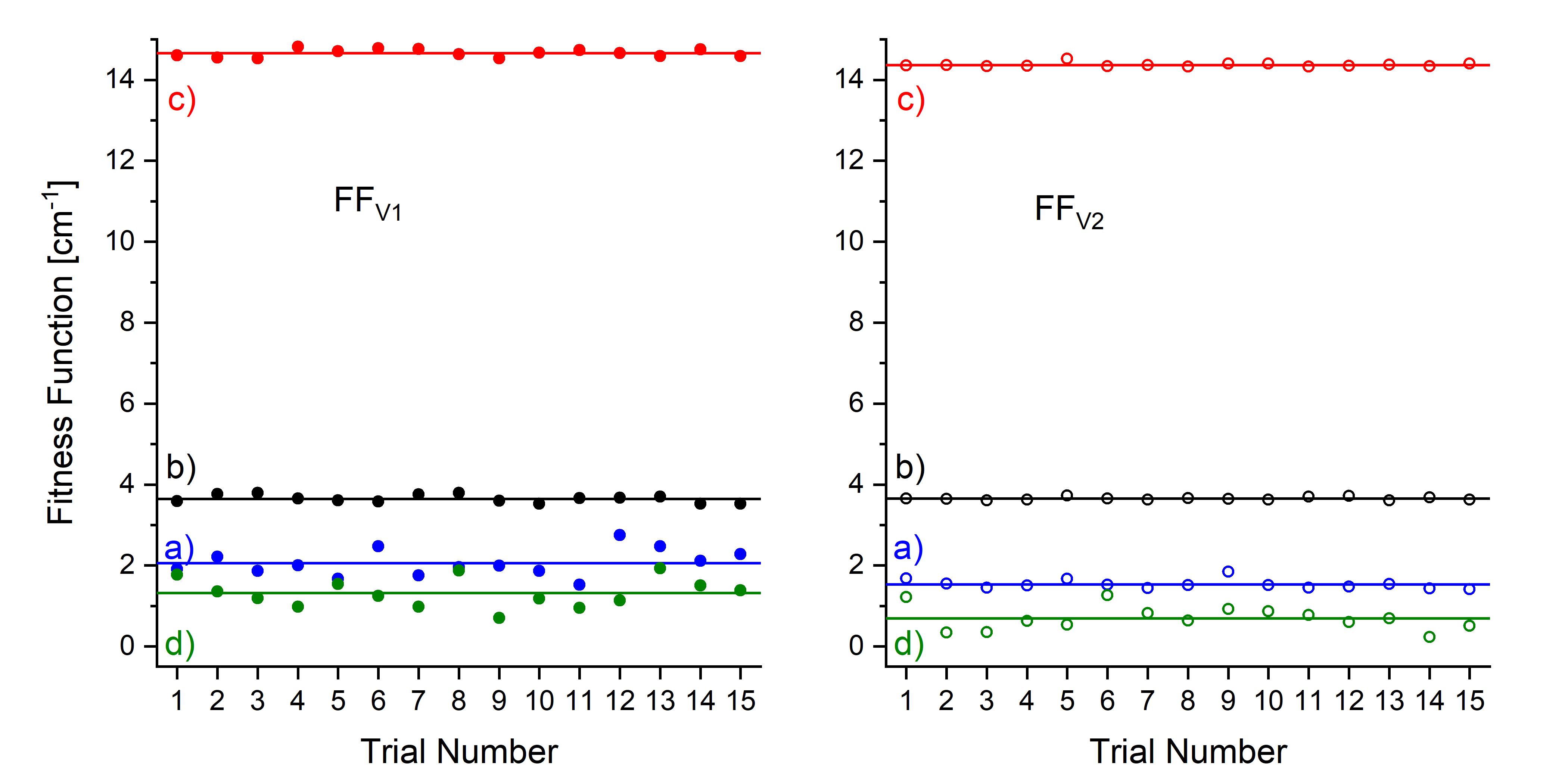}
	\caption{
		Fitness function ($FF$) obtained as a result of GA applied for "noised" data and two versions of $FF$: $FF_{V1}$ (left side) and $FF_{V2}$ (right side). (a), (b) and (c) Results for $FF_{V1}$ and $FF_{V2}$  for the best candidate solutions in independent trials for datasets with added Gaussian random noise, with $\sigma$ equal to 0.01 cm$^{-1}$, 0.025 cm$^{-1}$ and  0.1 cm$^{-1}$, respectively. (d) Result for $FF$ for a dataset without noise. The average values of the trials outcome in each test are shown with solid lines. Details in text.}		
	\label{figNoise}
\end{figure}

For both representations of $FF$, we performed three tests in which we added random noise with $\sigma$ equal to 0.01, 0.025 and 0.1 $cm^{-1}$. For each test, we run GA 15 times with the same parameters in each trial (10 generations, 800 candidate solutions in each generation, except the first generation with 2400 candidate solutions). $FF$ obtained in best solutions are presented in Fig. \ref{figNoise}, whereas Table \ref{tabNoise} collects potential parameters obtained using GA for each test, and the final result of a single test corresponds to the candidate solution with the lowest $FF$ obtained in 15 trials. As one can see, for both $FF$ representations the obtained potential parameters are in good agreement with expected ones, except value of $R_e$ that was derived for the strongest noise. From data presented in Table \ref{tabNoise} it is also evident, that for each test (and even for each trial in a single test), the value of $FF$ for the best candidate solution is comparable with the aggregated noise added to the dataset ($\Sigma_{\upsilon,J}|N|$). Due to the random nature of the noise, it is impossible to find an analytical potential that would provide $FF$ that is significantly smaller than the aggregated noise. Furthermore, to precisely simulate the "noised" spectrum one can try to use extremely complicated pointwise potential, but this result would be non-physical. The results of GA lead to values of $FF$s that are comparable to the aggregated noise, meaning GA can work correctly even in case of a "noised" datasets.

\begin{table}[bt!]
	\footnotesize
	\caption{Results of searching of EMO $(N=1)$ potential parameters using GA for data based on the potential inspired by the $b^30_u^+(5^3P_1)$ - state potential in Cd$_2$ \cite{Cd2IRPC} with added random noise to the energies of $(\upsilon,J)$ levels. Potential parameters are taken from the best candidate solution in the whole test (the result with lowest $FF$ among 15 trials).}
	\label{tabNoise}
	\begin{center}
		\begin{tabular}{ccccccc}
			\hline\\[-8pt]
			Parameter&Searching&GA&GA&GA&Expected\\ 
			&range&value$^a$&value$^b$&value$^c$&value$^d$\\ %
			\hline\\[-8pt]
			\multirow{2}{*}{$R_e$[\AA]}& 	\multirow{2}{*}{3.95-4.05} & 4.018$^e$  & 4.024$^e$  & 3.971$^e$ & 	\multirow{2}{*}{4.020} \\
			&  & 4.023$^f$ & 4.019$^f$  & 3.970$^f$ &  \\
			\multirow{2}{*}{$D_e[cm^{-1}]$}& 	\multirow{2}{*}{250-270} & 259.66$^e$ & 259.64$^e$ & 259.32$^e$ & 	\multirow{2}{*}{259.50} \\
			&  & 259.49$^f$ & 259.49$^f$ & 259.51$^f$ &  \\
			\multirow{2}{*}{$\beta_0$[1/\AA]}& 	\multirow{2}{*}{0.8-1.4} & 1.150$^e$ & 1.150$^e$ & 1.150$^e$ & 	\multirow{2}{*}{1.150} \\
			& & 1.150$^f$& 1.150$^f$ & 1.150$^f$ & \\
			\multirow{2}{*}{$\beta_1$[1/\AA]}& 	\multirow{2}{*}{0.0-0.6}& 0.174$^e$ & 0.173$^e$ & 0.190$^e$ & 	\multirow{2}{*}{0.180} \\
			& & 0.183$^f$& 0.181$^f$ & 0.178$^f$ & \\
			\hline\\[-8pt]
			\hline\\[-8pt]	
			${FF_{V1\,avg}}^g[cm^{-1}]$& &1.53(12) &3.65(4)& 14.37(5) &\\
			${FF_{V2\,avg}}^g[cm^{-1}]$& &2.06(33) &3.64(10) &14.66(10)  &\\
			
			\multirow{2}{*}{Sum of noise$^h$ [$cm^{-1}$]}& &1.37$^e$ &3.58$^e$ & 14.37$^e$ &	\\	
			& & 1.33$^f$ &3.58$^f$& 14.79$^f$ & \\
			\hline\\[-8pt]
		\end{tabular}	
	\end{center}    
	$^a$Added Gaussian random noise with $\sigma = $ 0.01 $cm^{-1}$.\\	
	$^b$Added Gaussian random noise with $\sigma = $ 0.025 $cm^{-1}$.\\
	$^c$Added Gaussian random noise with $\sigma = $ 0.1 $cm^{-1}$. \\		
	$^d$Parameters used for creation of the reference dataset.\\
	$^e$Results obtained for $FF_{V1}$\\
	$^f$Results obtained for $FF_{V2}$\\
	$^g$The average $FF$ for best solutions obtained in 15 trials (compare horizontal lines in Fig. \ref{figNoise}),  the uncertainty obtained as $\sigma$.\\
	$^h$Sum of noise added to $(\upsilon,J)$ levels: $\Sigma_{\upsilon,J}|N|$.\\	
\end{table}

\subsection{Performance of GA in case of missing data}
\label{missingData}
\begin{figure}[bt!]
	\centering
	\includegraphics[width=0.95\textwidth]{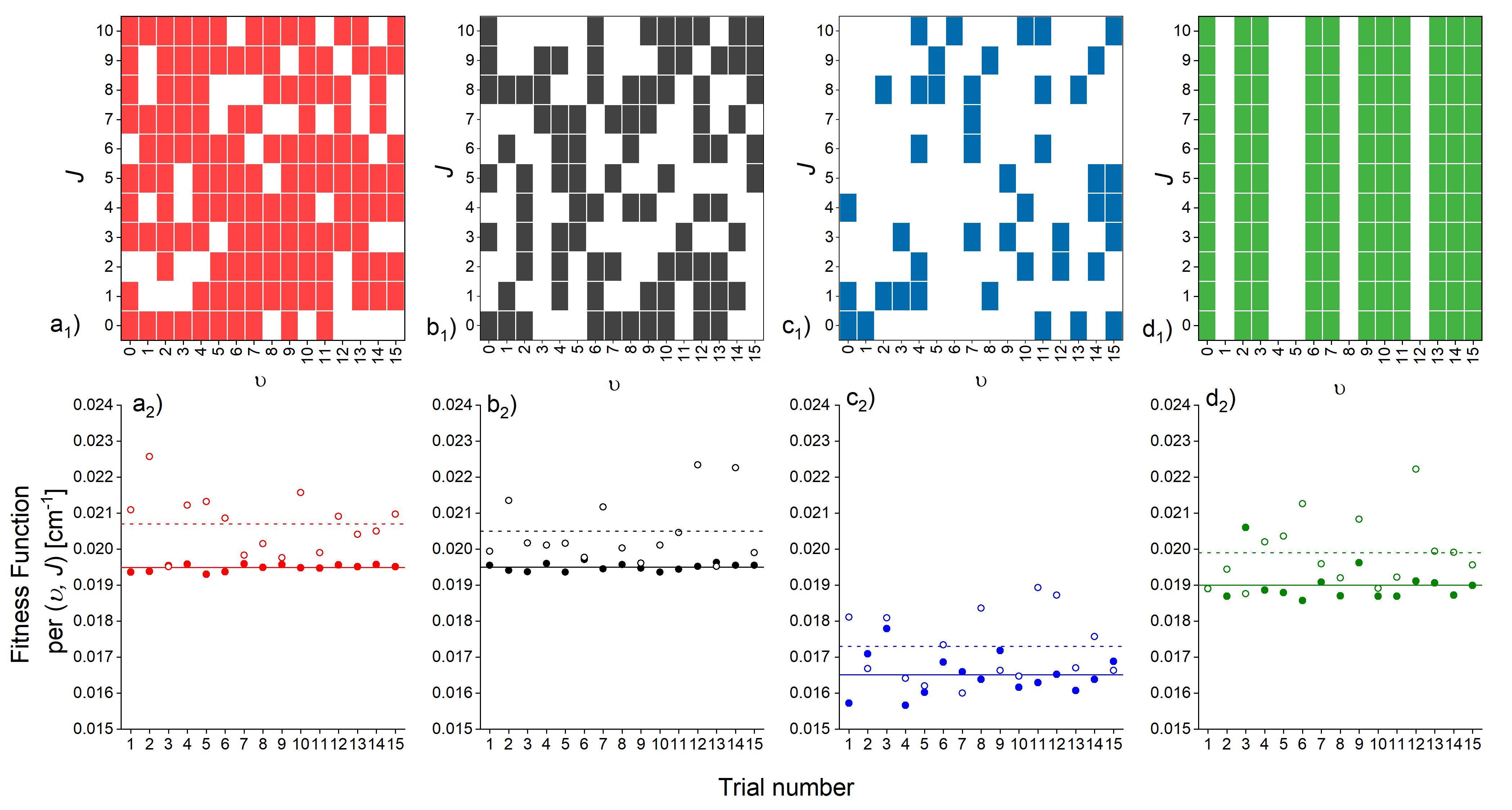}
	\caption{Fitness function ($FF$) obtained as a result of GA for four different cases of missing experimental data. Lower part: $FF$ per $(\upsilon, J)$ level in input data for the best candidate solutions obtained using GA in 15 trials for (randomly chosen) missing (a$_2$) 25\%, (b$_2$) 50\% and (c$_2$) 75\% $(\upsilon,J)$  levels . (d$_2$) Result for the case of missing of all of $\upsilon$=1, 4, 5, 8 and 12 vibrational components ($FF_{V1}$ and $FF_{V2}$ represented with full and empty circles, respectively). $FF_{V1}$ and $FF_{V2}$  averaged over 15 trials in each test are depicted with solid and dashed horizontal lines, respectively. Upper part: (a$_1$), (b$_1$), (c$_1$) and (d$_1$) present visualizations of missing $(\upsilon, J)$ levels  in the input data (white spaces) for simulations for which results are presented in (a$_2$), (b$_2$), (c$_2$) and (d$_2$), respectively. Details in text.}		
	\label{figLackLevels}
\end{figure}

To check the correctness of performance of GA in case when some of the experimental data are missing, we conducted four tests based on the artificially generated dataset for EMO $(N = 2)$ potential which initially contained 176 $(\upsilon,J)$ levels (from $\upsilon=0$ to $\upsilon=15$, and from $J=0$ to $J=10$). In the initial dataset, to $(\upsilon,J)$ levels we added random noise with $\sigma = 0.025$ cm$^{-1}$, as described in section \ref{noised}. For the first three tests (see Fig. \ref{figLackLevels} (a$_1$),(b$_1$) and (c$_1$)), from the considered dataset, we randomly eliminated 25\%, 50\% and 75\% of $(\upsilon,J)$ levels, respectively. For the fourth test (see Fig. \ref{figLackLevels} (d$_1$)), from the dataset we eliminated five vibrational components: $\upsilon$=1, 4, 5, 8 and 12. Entire elimination process is illustrated with white squares (Fig. \ref{figLackLevels}, upper part) that indicate which $(\upsilon,J)$ levels were eliminated in the respective tests.

From Eq. \ref{eqFF} and \ref{eqFFV2} one can see, that for both representations, value of $FF$ is associated with the number of $(\upsilon,J)$ levels in the dataset: the more levels in the dataset the higher value of $FF$. Thus, in order to easily compare the results of tests performed on the reduced datasets containing different number of $(\upsilon,J)$ levels, instead of $FF$ (Eq. \ref{eqFF} or \ref{eqFFV2}) we have to compare $FF$ per $(\upsilon,J)$ level \textit{i.e.}, $FF$ divided by the number of $(\upsilon,J)$ levels in the dataset.

Result of tests performed for both representations of $FF$ is presented in Fig. \ref{figLackLevels} (lower part). Plots in (a$_2$), (b$_2$) and (c$_2$) show the results corresponding to 25\%, 50\% and 75\% of $(\upsilon,J)$ levels missing in the dataset, respectively, while (d$_2$) points out the result for missing $\upsilon$=1, 4, 5, 8 and 12 entire vibrational components. Values of $FF_{V1}$ per $(\upsilon,J)$ level for the best candidate solutions obtained using GA in 15 independent trials are plotted with full circles, whereas empty circles represent values of $FF_{V2}$ for the best candidate solutions. $FF_{V1}$ and $FF_{V2}$  values averaged over results of all 15 trials in each test are shown with solid and dashed horizontal lines, respectively. Table \ref{tabLackOfLevels} collects potential parameters obtained for the best candidate solutions in each test (result from all 15 trials) for both representations of $FF$.

The obtained results show GA working sufficiently well when $(\upsilon,J)$ levels are missing in the analyzed spectrum. We can assume, that the correctness of the results obtained in each test can be described quantitatively by $FF$ per $(\upsilon,J)$ level ($FF_{PL\,avg}$) averaged over 15 trials in each test (compare with horizontal lines in Fig. \ref{figLackLevels}(a$_2$), (b$_2$), (c$_2$) and (d$_2$)), whereas a $\sigma$ can describe its uncertainty. From Table \ref{tabLackOfLevels} one can conclude that for both representations of $FF$, $FF_{PL\,avg}$ obtained for different sets with missing levels are comparable, only $FF_{PL\,avg}$ for dataset with missing 75\% levels are about 15\% lower as compared with results for other datasets.

\begin{table}[bt!]
	\footnotesize
	\caption{Results of searching of five EMO $(N=2)$ potential parameters using GA for data corresponding to $(\upsilon,J)$ levels missing from the spectrum. The dataset was based on a potential similar to the one for the $b^30_u^+(5^3P_1)$ state in Cd$_2$ \cite{Cd2IRPC}. Values of $R_e$, $D_e$ and $\beta_i$ are obtained in the entire test for the best candidate solution selected from results of 15 independent trials.}
	\label{tabLackOfLevels}
	\begin{center}
		\begin{tabular}{cccccccc}
			\hline\\[-8pt]
			Parameter&Searching&GA&GA&GA&GA&Expected\\ 
			&range&value$^a$&value$^b$&value$^c$&value$^d$&value$^e$\\ %
			\hline\\[-8pt]
			\multirow{2}{*}{$R_e$[\AA]} &\multirow{2}{*}{3.95-4.05}& 4.013$^f$ & 4.002$^f$ & 3.990$^f$ & 4.013$^f$ &\multirow{2}{*}{4.010} \\
			&& 3.997$^g$ & 3.998$^g$ & 3.986$^g$ & 3.996$^g$ & \\
			\multirow{2}{*}{$D_e[cm^{-1}]$} &\multirow{2}{*}{250-270}& 257.47$^f$ & 257.52$^f$ & 257.64$^f$ & 257.47$^f$ & \multirow{2}{*}{257.50} \\
			&& 257.49$^g$ & 257.49$^g$ & 257.49$^g$ & 257.50$^g$ & \\
			\multirow{2}{*}{$\beta_0$[1/\AA]}&\multirow{2}{*}{0.8-1.4} & 1.120$^f$ & 1.120$^f$ &1.120$^f$ & 1.120$^f$ & \multirow{2}{*}{1.120}\\
			&& 1.120$^g$ & 1.120$^g$ &1.120$^g$ & 1.120$^g$ &\\
			\multirow{2}{*}{$\beta_1$[1/\AA]}&\multirow{2}{*}{0-0.6}& 0.346$^f$ & 0.340$^f$ & 0.336$^f$ & 0.320$^f$ &\multirow{2}{*}{0.350} \\	
			&& 0.342$^g$ & 0.343$^g$ & 0.338$^g$ & 0.329$^g$ &\\
			\multirow{2}{*}{$\beta_2$[1/\AA]}&\multirow{2}{*}{0-0.25}& 0.077$^f$ & 0.083$^f$ &0.078$^f$ & 0.167$^f$ &\multirow{2}{*}{0.050}\\
			&& 0.085$^g$ &0.082$^g$ &0.096$^g$ &0.129$^g$ &\\
			\hline\\[-8pt]
			\hline\\[-8pt]
			\multirow{2}{*}{${FF_{PL\,avg}}^h[cm^{-1}]$}& &0.0195(1)$^f$&0.0195(1)$^f$&0.0165(5)$^f$&0.0190(5)$^f$\\	
			& &0.0207(8)$^g$&0.0204(9)$^g$&0.0172(10)$^g$&0.0199(10)$^g$\\	
			\hline\\[-8pt]
			
		\end{tabular}	
	\end{center}
	$^a$Obtained for 25\% of missing $(\upsilon,J)$ levels.\\
	$^b$Obtained for 50\% of missing $(\upsilon,J)$ levels.\\
	$^c$Obtained for 75\% of missing $(\upsilon,J)$ levels.\\
	$^a$Obtained for missing $\upsilon = 1, 4, 5, 8$ and $12$ vibrational components.\\
	$^e$Parameters used for creation of the reference dataset.\\
	$^f$Results obtained for  $FF_{V1}$.\\
	$^g$Results obtained for  $FF_{V2}$.\\
	$^hFF$ per ($\upsilon,J$) level averaged over 15 trials (compare with solid horizontal lines in Fig.  \ref{figLackLevels}, lower part), the uncertainty obtained as $\sigma$.\\	
\end{table}

\subsection{Efficiency of GA method}
To assess the efficiency of GA method, we compared its results with results of the simplest "brute force" method which is explained below. The most time consuming part of GA is associated with execution of LEVEL program which calculates energies of $(\upsilon,J)$ levels based on electronic state potential parameters.

Let us assume GA algorithm terminating after 10 generations and each generation contained 800 candidate solutions except the first generation which contains 2400 candidate solutions. Evaluation of $FF$ for one candidate solution is connected with a single execution of LEVEL, meaning that during the execution of GA, LEVEL is executed 9600 times. To assess a GA efficiency, we compared it with the simplest "brute force" algorithm which generated 9600 candidate solutions with random parameters chosen from specified ranges and returned a solution that has the lowest value of $FF$. We performed test for both $FF_{V1}$ and $FF_{V2}$. In the test, we used one of the "noised" dataset from Sec. \ref{noised} \textit{i.e.}, EMO $(N=1)$, and added Gaussian random noise with $\sigma$ = 0.025 cm$^{-1}$. Due to the same number of execution of LEVEL, GA and "brute force" algorithms have comparable execution times, which for workstation based on Intel(R) Xeon(TM) E3-1240 v3 processor with 32 GB RAM were about 12.7 minutes (average of 15 trials). Similar time - on average 10.5 minutes - was obtained for notebook with Intel(R) Core(TM) i7-8750H CPU and 16GB RAM.

\begin{figure}[bt!]
	\centering
	\includegraphics[width=0.8\textwidth]{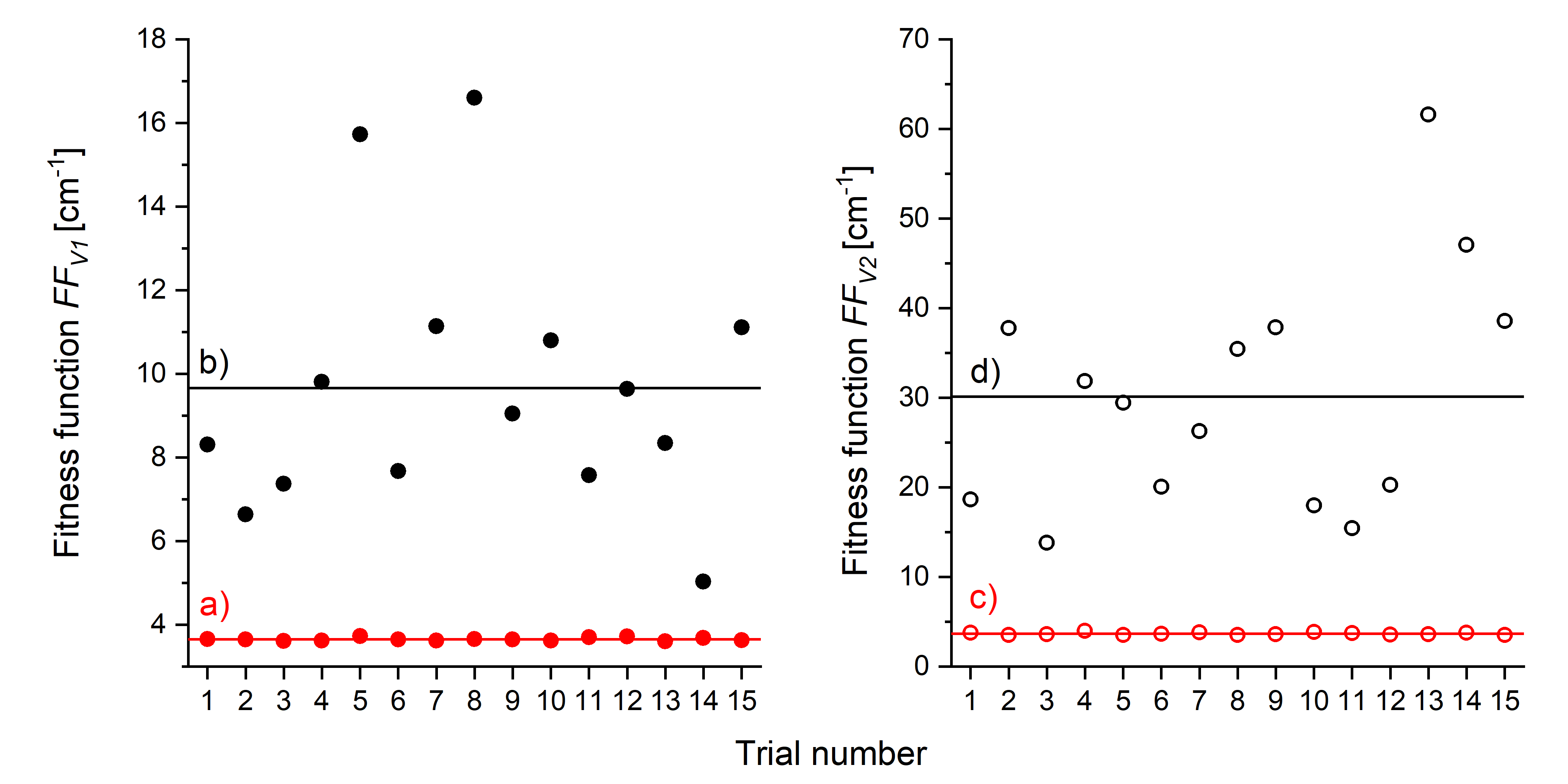}
	\caption{Comparison of fitness functions ($FF$) for best solution obtained using (a) and (c) GA (red circles), and (b) and (d) "brute force" algorithm (black circles) in 15 independent trials. Results for $FF_{V1}$ and $FF_{V2}$ are shown on the left (full circles) and the right (empty circles) parts, respectively. The mean values of 15 trials for GA and "brute force" algorithms are depicted with respective horizontal solid lines. Details in text.}		
	\label{figCompar}
\end{figure}

\begin{figure}[bt!]
	\centering
	\includegraphics[width=0.9\textwidth]{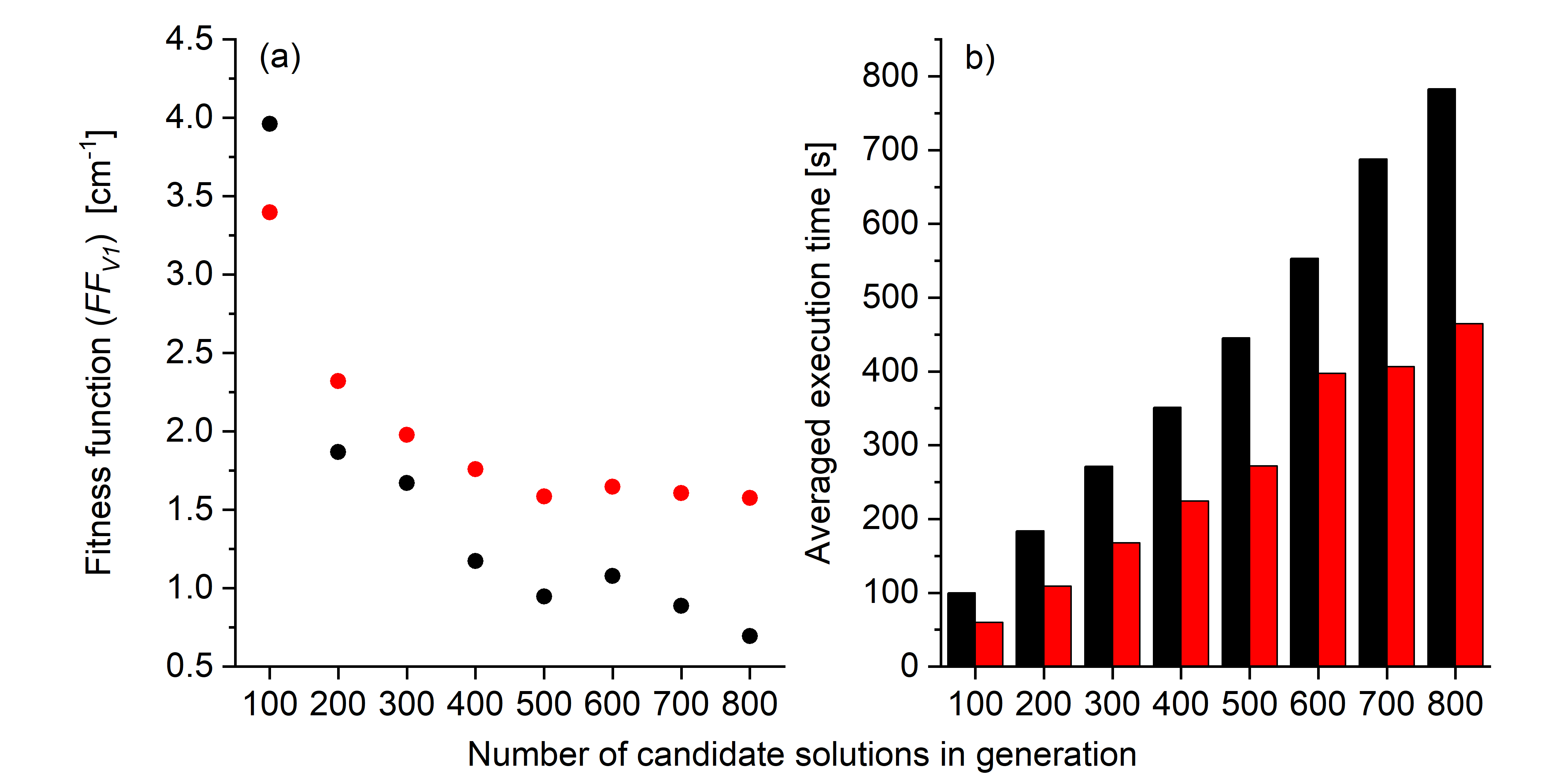}
	\caption{(a) Fitness function ($FF_{V1}$) and (b) averaged execution time of GA for different number of candidate solutions in each generation (except the first generation for which the number of solutions was tripled). Results averaged over 15 trials for GA that terminated after 5 (red circles and bars) and 10 (black circles and bars) generations. Details in text.}		
	\label{figTime}
\end{figure}

Fig. \ref{figCompar} presents comparison of GA with "brute force" algorithms for 15 independent trials, for both $FF$ representations. It is evident, that for $FF_{V1}$ and $FF_{V2}$, the results of GA have significantly lower values then results of "brute force" algorithm. The values of $FF_{V1}$ for the best solution averaged over 15 trials for GA and "brute force" algorithm are 9.6 cm$^{-1}$ and 3.7 cm$^{-1}$, respectively. Similar avarages computed for $FF_{V2}$ are equal to 3.7 and 30.15 cm$^{-1}$, respectively. Moreover, "brute force" algorithm has much higher dispersion than that of GA: 3.1 cm$^{-1}$ \textit{vs.} 0.038 cm$^{-1}$ for $FF_{V1}$ and 13 cm$^{-1}$ \textit{vs.} 0.15 cm$^{-1}$ for $FF_{V2}$.

To analyze performance and execution time of GA for different parameters, we run the algorithm for different number of candidate solutions in a generation (starting from 100 to 800 with 100 step) and for two different stop conditions (terminating GA after 5 and 10 generations). The result of tests for $FF_{V1}$ is presented in Fig. \ref{figTime} which shows the averaged $FF_{V1}$ (over 15 execution of GA) for the best candidate solution found by GA (left side), and the averaged execution time of GA executed using Intel(R) Xeon(TM) E3-1240 v3 with 32 GB RAM (right side).

\section{Tests of GA on experimental data}
\subsection{The $E^3\Sigma^+_{1\,in}(6^3S_1)$ state in CdAr and CdKr}

To check GA on real experimental data, we used it to find parameters of approximate analytical representation of the $E^3\Sigma^+_{1\,in}$ - state potential in CdAr and CdKr employing $FF_{V1}$ (see Eq. \ref{eqFFV2}). Experimental data \cite{URBANCZYK_CDAR,URBANCZYK_CdKr} as well as result of \textit{ab initio} calculations \cite{PhysRevA_CdAr} show, that the $E^3\Sigma^+_{1}$ state in both molecules has a double-well structure. Thus, to analyze its inner and outer potential wells IPA method is usually used.

\begin{table}[bt!]
	\footnotesize
	\caption{Experimental and simulated transition frequencies (in cm$^{-1}$) for the $E^3\Sigma_{1\,in}^+(6^3S_1),\upsilon'\leftarrow A^3\Pi_{0^+}(5^3P_1), \upsilon''=6$  transition in CdAr molecule.}
	\label{tabCdArEnergies}
	\begin{center}
		
		\begin{tabular}{cccccc}
			\hline\\[-8pt]
			$\upsilon'$&$\nu_{expt}^a$&$\nu_{sim}^b$&$\nu_{sim}^c$&$|\nu_{expt}-\nu_{sim}|^b$&$|\nu_{expt}-\nu_{sim}|^c$\\ 
			\hline\\[-8pt]
			1	&	19845.8	&	19845.8	&	19845.8	&	0.0	&	0.0			\\
			2	&	19944.7	&	19944.9	&	19944.2	&	0.2	&	0.5			\\
			3	&	20039.2	&	20039.8	&	20038.6	&	0.6	&	0.6			\\
			4	&	20129.4	&	20130.5	&	20129.1	&	1.0	&	0.3			\\
			5	&	20215.8	&	20216.9	&	20215.7	&	1.1	&	0.1			\\
			6	&	20298.1	&	20299.2	&	20298.1	&	1.1	&	0.0			\\
			7	&	20376.3	&	20377.3	&	20376.5	&	1.0	&	0.2			\\
			8	&	20450.4	&	20451.3	&	20450.8	&	0.9	&	0.4			\\
			9	&	20521.0	&	20521.0	&	20521.0	&	0.0	&	0.0			\\
			10	&	20586.5	&	20586.5	&	20586.9	&	0.0	&	0.4			\\
			11	&	20648.2	&	20647.8	&	20648.5	&	0.4	&	0.3			\\
			12	&	20705.8	&	20704.9	&	20705.9	&	0.9	&	0.1			\\
			13	&	20759.1	&	20757.9	&	20758.8	&	1.2	&	0.3			\\
			14	&	20807.9	&	20806.6	&	20807.4	&	1.3	&	0.5			\\
			15	&	20852.1	&	20851.2	&	20851.4	&	0.9	&	0.7			\\
			16	&	20891.5	&	20891.5	&	20890.8	&	0.0	&	0.7			\\
			17	&	20925.4	&	20927.7	&	20925.5	&	2.3	&	0.1			\\
			18	&	20952.6	&	20959.6	&	20955.4	&	7.0	&	2.8			\\
			\hline\\[-8pt]
			\multicolumn{4}{l}{Sum}&20.0&8.2\\	    
			\hline\\[-8pt]
		\end{tabular}	
	\end{center}
	$^a$ Experimental values \cite{URBANCZYK_CDAR}.\\	
	$^b$ Simulation based on a  Morse representation of the $E^3\Sigma_{1\,in}^+$ potential; parameters obtained using GA for $\upsilon'$ from 1 to 18.\\
	$^c$ Simulation based on an EMO representation of the $E^3\Sigma_{1\,in}^+$ potential; parameters obtained using GA for $\upsilon'$ from 1 to 18.
\end{table} 

\begin{table}[h!]
	\footnotesize
	\caption{Experimental and simulated energies of vibrational levels (in cm$^{-1}$)  for the $E^3\Sigma_{1\,in}^+(5^3P_1)$ state in CdKr molecule. }
	\label{tabCdKrEnergies}
	\begin{center}
		\begin{tabular}{cccccc}
			\hline\\[-8pt]
			$\upsilon'$&$\nu_{expt}^a$&$\nu_{sim}^b$&$\nu_{sim}^c$&$|\nu_{expt}-\nu_{sim}|^b$&$|\nu_{expt}-\nu_{sim}|^c$\\ 
			\hline\\[-8pt]
			
			0	&	49909.1	&	49909.1	&	49909.1	&	0.0	&	0.0	\\
			1	&	49997.2	&	49997.0	&	49996.3	&	0.2	&	0.9	\\
			2	&	50082.6	&	50082.4	&	50081.3	&	0.2	&	1.3	\\
			3	&	50163.3	&	50165.3	&	50163.9	&	2.0	&	0.6	\\
			4	&	50243.8	&	50245.8	&	50244.1	&	2.0	&	0.3	\\
			5	&	50322.3	&	50323.7	&	50322.0	&	1.4	&	0.3	\\
			6	&	50398.5	&	50399.2	&	50397.6	&	0.7	&	0.9	\\
			7	&	50470.7	&	50472.2	&	50470.7	&	1.5	&	0.0	\\
			8	&	50542.3	&	50542.7	&	50541.5	&	0.4	&	0.8	\\
			9	&	50610.3	&	50610.8	&	50609.8	&	0.5	&	0.5	\\
			10	&	50676	&	50676.4	&	50675.7	&	0.4	&	0.3	\\
			11	&	50738.6	&	50739.4	&	50739.1	&	0.8	&	0.5	\\
			12	&	50800	&	50800.0	&	50800.1	&	0.0	&	0.1	\\
			13	&	50858.4	&	50858.2	&	50858.6	&	0.2	&	0.2	\\
			14	&	50914.4	&	50913.8	&	50914.6	&	0.6	&	0.2	\\
			15	&	50967.9	&	50967.0	&	50968.1	&	0.9	&	0.2	\\
			16	&	51018.8	&	51017.7	&	51019.0	&	1.1	&	0.2	\\
			17	&	51067.3	&	51065.9	&	51067.3	&	1.4	&	0.0	\\
			18	&	51113.2	&	51111.6	&	51113.1	&	1.6	&	0.1	\\
			19	&	51156.5	&	51154.9	&	51156.2	&	1.6	&	0.3	\\
			20	&	51197.2	&	51195.7	&	51196.7	&	1.5	&	0.5	\\
			21	&	51235	&	51234.0	&	51234.5	&	1.0	&	0.5	\\
			22	&	51269.7	&	51269.8	&	51269.6	&	0.1	&	0.1	\\
			23	&	51302.1	&	51303.1	&	51302.0	&	1.0	&	0.1	\\
			24	&	51331.3	&	51334.0	&	51331.5	&	2.7	&	0.2	\\
			25	&	51357.7	&	51362.3	&	51358.3	&	4.6	&	0.6	\\

			\hline\\[-8pt]
			\multicolumn{4}{l}{Sum}&28.5&10.0\\
			
			\hline\\[-8pt]
		\end{tabular}	
	\end{center}
	$^a$ Experimental values \cite{URBANCZYK_CdKr}.\\	
	$^b$ Simulation based on a Morse representation of the $E^3\Sigma_{1\,in}^+$ potential; parameters obtained using GA for $\upsilon'$ from 0 to 25\\
	$^c$ Simulation based on an EMO representation of the  $E^3\Sigma_{1\,in}^+$ potential; parameters obtained using GA for $\upsilon'$ from 0 to 25.
\end{table}

\begin{figure}[bt!]
	\centering
	\includegraphics[width=0.85\textwidth]{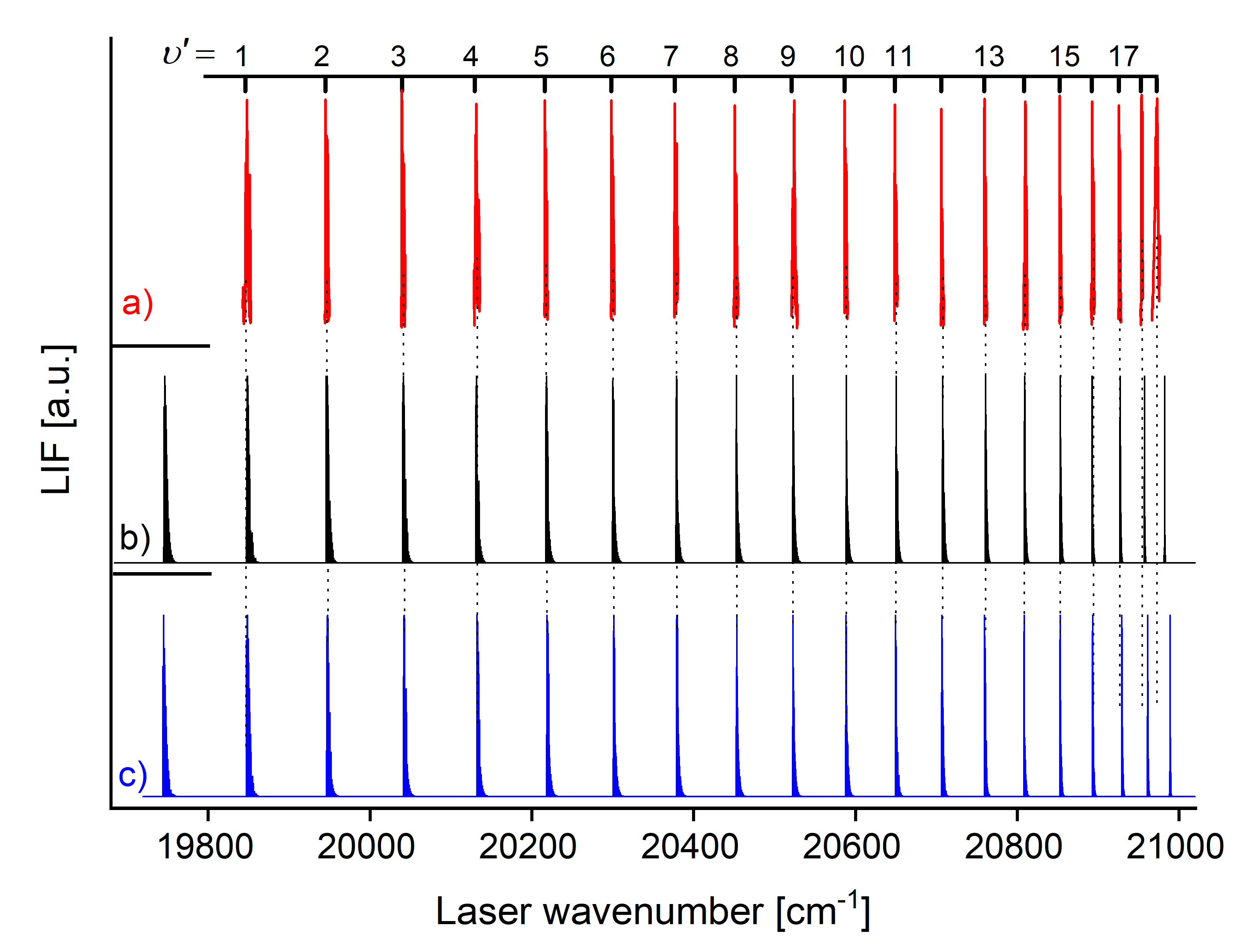}
	\caption{(a) Experimental LIF excitation spectrum of the  $E^3\Sigma_{1\,in}^+(6^3S_1),\upsilon'\leftarrow A^3\Pi_{0^+}(5^3P_1),\upsilon''=6$ transition in CdAr. Simulations performed using (b) EMO $(N=1)$ and (c) Morse representation of the $E^3\Sigma_{1\,in}^+$ - state potential obtained using GA. Intensities of vibrational components in the experimental spectrum and in both simulations were normalized.}
	\label{fig2}
\end{figure}

\begin{table}[h!]
	\footnotesize	
	\caption{
		Parameters of Morse and EMO representations of the $E^3\Sigma_{1\,in}^+(6^3S_1)$ - state potential in CdAr and CdKr obtained by GA.}
	\label{tabCdArParam}
	\begin{center}		
		\begin{tabular}{cccc|ccc}   
			\hline\\[-8pt]
			&\multicolumn{3}{c}{CdAr}&\multicolumn{3}{c}{CdKr}\\
			\hline\\[-8pt]
			Parameter &Searching &GA &GA &Searching &GA &GA \\ 
			&range &value$^a$ &value$^b$ &range &value$^a$ &value$^b$ \\ %
			\hline\\[-8pt]
			$R_e$[\AA] & 2.85-2.85$^c$ & 2.85 & 2.85 & 2.99-2.99$^c$ &2.99 &2.99 \\
			$D_e[cm^{-1}]$ & 1330-1390 & 1376.64 & 1337.34 &1550-1700& 1646.61 & 1599.89  \\
			$\beta_0$[1/\AA] & 1.8-2.1 & 1.919 &1.921 &1.7-2.2& 1.885& 1.895   \\
			$\beta_1$[1/\AA] & 0.0-0.8& - &  0.743 &0.0-1.0& - &0.712  \\	
			\hline	
		\end{tabular}	
	\end{center}
	$^a$ Result for a Morse representation.\\
	$^b$ Result for an EMO $(N=1)$ representation.\\
	$^c$ Due to the fact that the experimental spectra do not reveal a resolved rotational structure, values of $R_e$ are fixed as their impact on the energies of vibrational components is negligible.		
\end{table}

Table \ref{tabCdArEnergies}  (second column) presents the $E^3\Sigma_{1\,in}^+,\upsilon'\leftarrow A^3\Pi_{0^+}, \upsilon''=6$ transition frequencies in CdAr recorded in OODR experiment \cite{URBANCZYK_CDAR}, whereas Table \ref{tabCdKrEnergies} collects experimental energies of vibrational levels of the $E^3\Sigma_{1\,in}^+$ state in CdKr \cite{URBANCZYK_CdKr}. For both molecules, we used GA to find parameters of two simple analytical representations of the $E^3\Sigma_{1\,in}^+$ - state potential: Morse and EMO $(N=1)$. In each case, GA terminated after 10 generations, each generation had 400 candidate solutions (except first generations which had 1200 candidate solutions). The obtained results are presented in Table \ref{tabCdArParam}. The observed differences between values of $D_e$ for Morse and EMO representations are associated with the fact, that both representations should correlate to different asymptotes. Due to the fact that for CdAr as well as for CdKr the $E^3\Sigma_1^+$ state has a potential barrier, both Morse and EMO representations are not suitable representations of the real potential near the dissociation limit, so they do not correlate to the atomic Cd asymptote. The asymptotes of both potentials should be chosen to obtain proper simulations of absolute energies (similar approach was used previously \textit{e.g.}, in \cite{URBANCZYK_CDAR}).

Fig. \ref{fig2} presents experimental spectrum (trace a) and its simulations based on an EMO (N=1) (trace b) and a Morse (trace c) representations of the $E^3\Sigma_{1\,in}^+$ - state potential in CdAr. Both simulations were obtained using PGOPHER program \cite{PGOPHER}. One can see, that for the $E^3\Sigma_{1\,in}^+$ state in CdAr (as well as that in CdKr), the simplest version of an EMO representation leads to significantly better simulation as compared with this based on a Morse representation: The sum of absolute values of discrepancies between simulated and measured energies of vibrational components was reduced from 20.0 cm$^{-1}$  to 8.2 cm$^{-1}$  for CdAr and from 28.5 cm$^{-1}$  to 10 cm$^{-1}$  for CdKr. Tests show, that including additional terms ($\beta_i$ for $i\geq2$) does not lead to significant improvement of the simulation (\textit{e.g.}, for CdAr, including $\beta_2$ in GA leads to a decrease of the sum of discrepancies from 8.2 to 7.1 cm$^{-1}$). Moreover, as we do not want to find the most accurate analytical potential to simulate observed spectra, it is justified of using a simpler version of EMO.
Our goal was to find a method of finding parameters of a simple analytical potential, which lead to a better simulation of experimentally observed energies of ro-vibrational levels than that offered by a Morse potential which can be used as a starting potential in IPA method. 


\section{Conclusions}

We employed a simple Genetic Algorithm (GA) to fit parameters of an analytical potential to spectroscopic data. We propose two representations of fitness function $FF$: $FF_{V1}$ and $FF_{V2}$, the former using differences between energies of $(\upsilon,J)$ levels, and the latter using these energies directly. Obtained analytical potential representations can be used as so-called starting potentials in the inverse perturbation approach (IPA) method. To check the correctness of GA, we tested it on the artificially generated reference data, which based on potentials with known parameters. Tests show that GA can precisely determine parameters of an expanded Morse oscillator (EMO) potential \cite{EMO1} with 4 or 5 parameters ($N=1$ and $N=2$, respectively). The algorithm can also run properly for "noised" experimental data and in case of missing $(\upsilon,J)$  levels that are eliminated from the analysis (see Sec. \ref{noised} and \ref{missingData}, respectively).

We also used GA to find parameters of an EMO function for the $E^3\Sigma_{1\,in}^+$ - state potential in CdAr and CdKr, based on the experimental spectra \cite{URBANCZYK_CDAR,URBANCZYK_CdKr} recorded with vibrational resolution. The energies of vibrational levels associated with the obtained EMO potentials were significantly closer to the experimental ones than those given by Morse potentials: Resulting sum of total discrepancies equal to 8.2 cm$^{-1}$ instead of 20.0 cm$^{-1}$ and 10.0 cm$^{-1}$ instead of 28.5 cm$^{-1}$ for CdAr and CdKr, respectively (for details see Tables \ref{tabCdArEnergies} and \ref{tabCdKrEnergies}). Results show, that GA can be used to obtain the starting potential for IPA method. Observed reduction in discrepancies between simulation based on the starting potential and experimental energies should simplify application of IPA method. GA can work with any analytical potential and we showed result for EMO potential as an example.

\section*{Acknowledgements}
This work was supported by the National Science Centre Poland under grant number UMO-2015/17/B/ST4/04016.

\bibliographystyle{tfnlm}
\bibliography{manuscript}

\end{document}